\documentclass[twocolumn]{IEEEtran}
\usepackage[T1]{fontenc}
\usepackage[latin9]{inputenc}
\usepackage{color}
\usepackage{array}
\usepackage{graphicx}
\usepackage{tabularx}
\usepackage{makecell}
\usepackage{cite}
\usepackage{changes}

\usepackage[unicode=true,
 bookmarks=true,bookmarksnumbered=true,bookmarksopen=true,bookmarksopenlevel=1,
 breaklinks=false,pdfborder={0 0 0},pdfborderstyle={},backref=false,colorlinks=false]
 {hyperref}
\hypersetup{pdftitle={Your Title},
 pdfauthor={Your Name},
 pdfpagelayout=OneColumn,pdfnewwindow=true, pdfstartview=XYZ, plainpages=false}

\makeatletter

\usepackage[caption=false,font=footnotesize]{subfig}
\usepackage{algorithm}
\usepackage{algorithmic}

\usepackage{multirow} 
\usepackage{amsmath} 
\usepackage{xcolor}

\allowdisplaybreaks[4]

\ifCLASSOPTIONcompsoc
\usepackage[caption=false,font=normalsize,labelfont=sf,textfont=sf]{subfig}
\else
\usepackage[caption=false,font=footnotesize]{subfig}
\fi

\usepackage{cite}
\usepackage{bm}
\usepackage{algorithmic}
\usepackage{algorithm}
\usepackage{graphicx}
\IEEEoverridecommandlockouts

\usepackage{lettrine}

\makeatother

\begin{document}
\title{\textcolor{black}{ Hierarchical Observe-Orient-Decide-Act Enabled UAV Swarms in Uncertain
Environments: Frameworks, Potentials, and Challenges}\textcolor{blue}{\index{}}}
\author{Ziye Jia, \IEEEmembership{Member,~IEEE}, Yao Wu, Qihui Wu, \IEEEmembership{Fellow,~IEEE}, Lijun He, \IEEEmembership{Member,~IEEE}, \\Qiuming Zhu, \IEEEmembership{Senior Member,~IEEE}, Fuhui Zhou, \IEEEmembership{Senior Member,~IEEE},
 and Zhu Han, \IEEEmembership{Fellow,~IEEE}

\thanks{This work was supported by National Natural Science Foundation of China under Grant 62231015 and 62301251. and in part by the Postgraduate Research \& Practice Innovation Program of Jiangsu Province under Grant SJCX25\_0152. (\textit{Corresponding author: Yao Wu}).}

\thanks{Ziye Jia, Yao Wu, Qihui Wu, Qiuming Zhu, and Fuhui Zhou are with the College of Electronic
and Information Engineering, Nanjing University of Aeronautics and
Astronautics, Nanjing 211106, China, (e-mail: jiaziye@nuaa.edu.cn, wu\_yao@nuaa.edu.cn, wuqihui@nuaa.edu.cn,
zhuqiuming@nuaa.edu.cn, zhoufuhui@nuaa.edu.cn).

Lijun He is with the School of Information and Control Engineering, China University of Mining and Technology, Xuzhou 221116, China, (e-mail: lijunhe.xd@gmail.com).

Zhu Han is with the University of Houston, TX 77004, USA, and also with the Department of Computer Science and
Engineering, Kyung Hee University, Seoul, 446-701, South Korea (hanzhu22@gmail.com ).

\textit{\textcolor{black}{}}}}
\maketitle
\begin{abstract}
Unmanned aerial vehicle (UAV) swarms are increasingly explored for their potentials in various applications such as surveillance, disaster response, and military.  
However, UAV swarms face significant challenges of implementing effective and rapid decisions under dynamic and uncertain environments. 
The traditional decision-making frameworks, mainly  relying on centralized control and rigid architectures, are limited by their adaptability and scalability especially in complex environments. 
To overcome these challenges, in this paper, we propose  a hierarchical Observe-Orient-Decide-Act (H-OODA) loop based framework for the UAV swarm operation in uncertain environments, which is implemented by embedding the classical OODA loop across the cloud-edge-terminal layers, and leveraging the network function virtualization (NFV) technology to provide flexible and scalable decision-making functions.
In addition, based on the proposed H-OODA framework, we joint autonomous decision-making and cooperative control to enhance the adaptability and efficiency of UAV swarms. 
Furthermore, we present some typical case studies to verify the improvement and efficiency of the proposed framework. 
Finally, the potential challenges and possible directions are analyzed to provide insights for the future H-OODA enabled UAV swarms. 
\end{abstract}


\section{Introduction}

Unmanned aerial vehicles (UAVs), composed of  multiple coordinated drones, are characterized by their flexibilities and  applied in various domains, such as  surveillance, search, rescue and agriculture,  especially in areas inaccessible to humans \cite{harchaoui20245g}, \cite{motlagh2017uav}.
Besides, due to the potentials to operate collaboratively, autonomously, and adaptively in complex environments, UAV swarms gain significant attentions from both industries and academies \cite{wu2022unmanned}. 
However, the uncertain environments may disrupt the operations of UAV swarms due to the severe weather conditions, unexpected obstacles, etc. These challenges highlight the requirements for robust control frameworks and advanced decision-making mechanisms that can enable UAV swarms to adapt and respond to varying  environmental conditions \cite{ouahouah2021deep}. In particular, the ability to make informed decisions in real-time is significant for UAV swarms to operate effectively confronting with uncertain environments.

To address the complexity and challenges associated with UAV swarms, the integration of advanced decision-making frameworks and network management technologies becomes  necessary \cite{bekkouche2020service}. To this end, the Observe-Orient-Decide-Act (OODA) decision-making framework, originally developed for military applications, is widely adopted in various domains due to its ability to enhance the autonomy and responsiveness in dynamic and unpredictable situations. By leveraging OODA, UAV swarms can improve their responsiveness to  environmental variations and make informed decisions in real-time. 
Meanwhile, the network function virtualization (NFV) technology enables the creation of virtualized network functions, which promote the flexibility, scalability, and efficiency in network management. Therefore, the integration of OODA and NFV provides UAV swarms with enhanced decision-making capabilities and optimized network resource allocation,  enabling  the effective operations in complex and uncertain environments. 

\begin{figure*}[ht]
\centering
\includegraphics[scale=0.6]{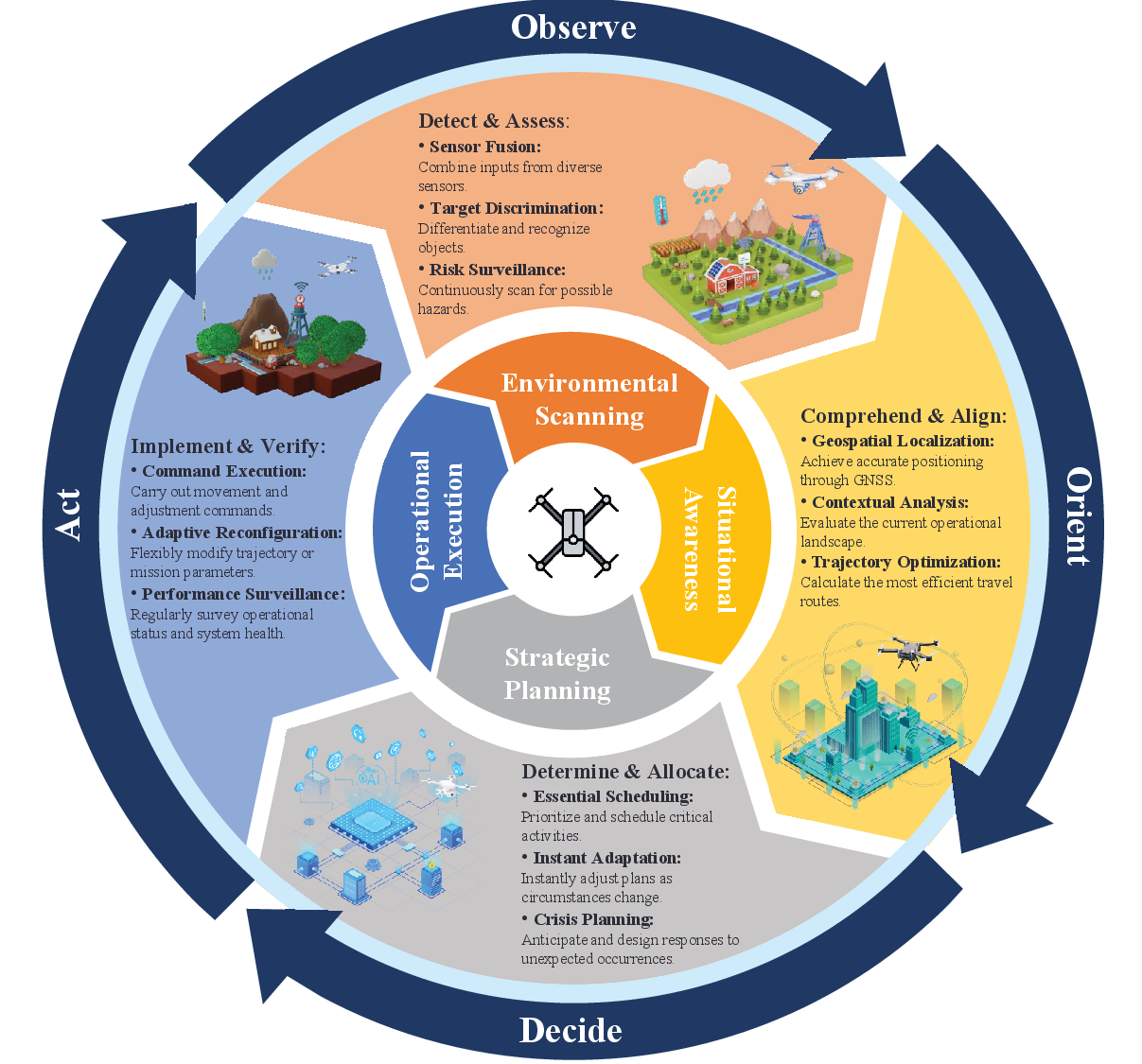}
\caption{\label{fig:Single-OODA}Illustration of the classic OODA loop for a single UAV.}
\end{figure*}

There exist some recent researches on the combination of OODA and UAVs, exploring the potentials to enhance the strategic decision-making and collaborative mission effectiveness. 
For example,  \cite{sun2023effectiveness} proposed an approach to evaluate the effectiveness of reconnaissance and strike UAV formations based on OODA, while another work \cite{alharasees2023evaluating} integrated OODA with artificial intelligence (AI) in UAV systems to provide a structured approach for decision-making.
Although these studies  have highlighted the potential of OODA in improving strategic decision-making for UAVs, the existing OODA-based methods typically focus on specific aspects of UAV operations, without considering the complexities of multi-UAV systems.
Moreover, \cite{qu2023elastic} proposed an architecture for elastic collaborative edge intelligence in UAV swarms, incorporating OODA principles to improve resource allocation decisions, and \cite{chen2023improving} incorporated OODA to improve the physical layer security in multi-UAV systems against hybrid wireless attacks. 
While these researches have applied the OODA principles to multi-UAV systems, the related  approaches basically rely on traditional architectures, which are limited in terms of scalability and flexibility. 
Therefore, it is necessary to explore  a more comprehensive and adaptable framework to effectively integrate multiple UAVs and enable efficient decision-making in complex and dynamic environments \cite{bouzid20235g}.  

Hence, building upon the Cloud-Edge-Terminal (CET) architecture, we propose the hierarchical OODA (H-OODA) framework to facilitate collaborative decision-making and adaptive management of UAV swarms. 
In particular, H-OODA can facilitate efficient management of UAV swarms via cooperative perception, decision, execution, and reconfiguration. 
We firstly illustrate the detailed layers of the H-OODA framework, as well as the detailed mechanisms in both the intra-network and inter-network of UAV swarms. Then, we analyze the integration of hierarchical OODA and NFV for efficient UAV swarm operations. 
Besides, we provide case studies to verify the performance of the proposed H-OODA framework in UAV swarms.  
Moreover, the challenges and future research directions are discussed.
In short, the key contributions of this work are summarized as follows.

\begin{itemize}
\item We develop a novel CET H-OODA framework integrating cloud, edge, and terminal layers to improve autonomous decision-making, data processing, and action coordination for UAV swarms in dynamic environments.
\item The NFV technology is innovatively incorporated into the  H-OODA framework to enable flexible and scalable deployment of network functions across different layers in UAV swarms, thereby enhancing the overall efficiency and adaptability. 
\item We design the hierarchical decision-making mechanism that leverages a layered architecture to fuse local and global information, enabling more informed and timely decision-making.
\end{itemize}

\section{H-OODA Framework for UAV Swarms}
\subsection{Overview of Classical OODA Loop }

The classical OODA loop, originally  proposed by U.S. Air Force Colonel John Boyd, serves as a foundational decision-making framework adopted across various domains \cite{OODA_Boyd2020}, particularly in volatile, uncertain, and complex scenarios.
Fig. \ref{fig:Single-OODA} illustrates the fundamental four-stage interactions between the agents and environments, which specifically depicts the application of the OODA loop for a single UAV, highlighting the practical implementation of the framework. 
The theoretical basis of the OODA loop stems from cognitive psychology and information processing theories, emphasizing the significance of situational awareness and decision-making speed.  
Besides, the efficacy of the loop lies in facilitating rapid decision-making in complex and dynamic environments by breaking down the  process into manageable stages. 
Fig. \ref{fig:Single-OODA} outlines the procedures and functions of a UAV at each stage of the OODA loop, detailed as follows.

\begin{figure*}[t]
\centering
\includegraphics[scale=0.75]{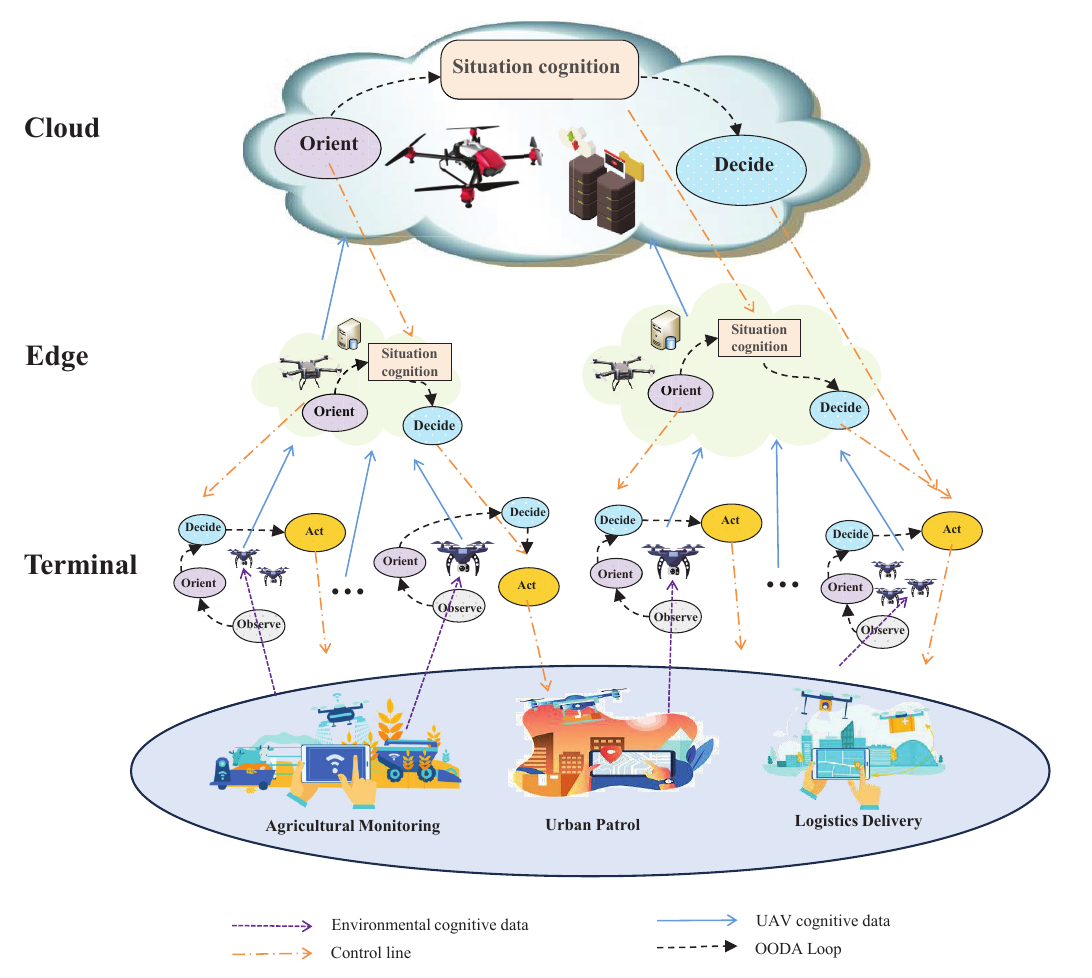}
\caption{\label{fig:OODA}Nested CET H-OODA framework within UAV swarms.}
\end{figure*}

\begin{itemize} 
\item \textit{Observe:} Initially, the environment is observed both actively and passively using sensor fusion techniques to detect and assess risks, including target discrimination and continuous risk surveillance.
\item \textit{Orient:} During the orientation, the UAV achieves geospatial localizations through global navigation satellite systems (GNSS)  for accurate positioning, and conducts contextual analysis to evaluate  current operational scenarios, along with  possible trajectories optimization. 
\item \textit{Decide:} In this stage, through internal or online discussions and inferences, the UAV determines the scheduling and instant adaptation plans, considering all potential outcomes.
\item \textit{Act:} The final stage is executing the decision from the previous stage, adjusting actions as needed during implementation to ensure effectiveness and alignment with desired outcomes. \end{itemize}

The classical OODA loop is a cyclical process where the outcome of the \textit{Act} stage becomes the new input for the \textit{Observe} stage, creating a continuous feedback. 
This iterative process enables  a UAV to learn from its experiences and adjust the strategies, and accordingly adapt to  circumstance variations and improve the decision-making. 
Consequently, UAVs can respond promptly to emerging situations while minimizing error risks.
Moreover, the four stages of the OODA loop can overlap to a certain extent, since they are not distinct and separate exactly. 
For example, the \textit{Orient} stage may involve some degree of decision-making, while the \textit{Decide} stage can necessitate additional observation and analysis. Such flexibility is vital in real-world scenarios where stage boundaries are generally indistinct.

Due to these characteristics,  the classical OODA loop for a UAV has  been widely applied in diverse domains such as military and business. However, as for the  UAV swarms in complex environments, a single OODA loop is not applicable directly.

\subsection{H-OODA Design for UAV Swarms}

\subsubsection{CET Architecture}

The operational effectiveness of UAV swarms in complex environments such as logistics delivery, urban patrol, and agricultural monitoring, relies on the ability to process vast amounts of data, expedite decision-making, and coordinate actions harmoniously. 
To achieve this goal, we propose the  CET H-OODA framework, as shown in Fig. \ref{fig:OODA}, which leverages the advantages of  stratified organizational design, enabling decentralized decision-making while maintaining overall coordination. 
To clarify the efficacy of the CET H-OODA framework, we analyze the critical role of its hierarchical structure in enhancing spectrum situation awareness, which is significant for the effective operation of UAV swarms. 
Note that the CET H-OODA structure primarily consists of the terminal layer, edge layer, as well as cloud layer, elaborated as follows.

\begin{itemize} \item \textit{Terminal Layer}: Multiple UAV swarms comprise the terminal layer, with each swarm consisting of a group of UAVs that collaborate to form a local frequency spectrum situation awareness map. The OODA loop at each swarm processes local data to inform decisions on frequency usage, facilitating adaptive transmission parameter adjustments in response to  local condition variations.

\item \textit{Edge Layer}: The edge layer fuses data from multiple terminal UAV swarms, forming a regional frequency spectrum situation awareness map. By integrating data from multiple terminals, the OODA loop at the edge layer enables regional frequency usage decisions, allowing for adaptive transmission parameter adjustments in response to varying regional conditions.

\item \textit{Cloud Layer}: The cloud layer provides a global spectrum situation awareness map by integrating data from multiple edge layers, incorporating 3D terrain mapping and dynamic interference modeling for comprehensive mission planning. The OODA loop at the cloud layer facilitates global frequency usage decisions, enabling adaptive transmission parameter adjustments in response to the global condition variation. While batch processing of edge-layer data inherently introduces latency, the lightweight predictive caching strategies  can significantly reduce synchronization delays without compromising global decision accuracy \cite{ameur2024efficient}. \end{itemize}

With the CET layers, H-OODA enables large-scale data processing and decision-making. 
For instance, in the logistics delivery, the hierarchical approach allows UAVs to gather and analyze real-time spectrum data, ensuring reliable communication with the control center. 
As for urban patrols, the hierarchical design supports dynamic adaptation of surveillance strategies based on continuous spectrum situation awareness, enabling drones to optimize their operations in response to varying frequency conditions. 
In agricultural monitoring,  H-OODA  facilitates the efficient transmission and analysis of crop health data, allowing UAVs to adjust their monitoring strategies based on real-time spectrum states. 
By coordinating multiple UAV swarms within this  framework,  H-OODA enhances the situational awareness and operational effectiveness, demonstrating the synergy between the hierarchical approach and spectrum situation awareness.

\subsubsection{Nested H-OODA Loops}

The CET architecture facilitates UAV swarms leveraging the OODA loop at each layer. However, the seamless coordination and decision-making require collaborative interactions among the OODA loops at different levels. 
To address this point, the H-OODA loops are nested to enable information exchange and synchronization cross CET-layers, characterized by embedding the OODA loops at each level within adjacent levels. 
Additionally, the nested structure allows each layer to maintain a consistent view of the operating environment while adapting to changing conditions in a unified and coordinated manner.

For instance, when a terminal UAV swarm detects  any abnormity about the local  situation, the OODA loop at the  swarm rapidly assesses the situation and initiates a response. 
The message is then transmitted to the edge OODA loops, which integrates all the information from  terminal UAV swarms to form a regional understanding of the current situation. 
Accordingly, the edge OODA loop  subsequently adjusts its decision-making parameters to account for the varying regional conditions and transmits these adjustments to both the terminal and cloud OODA loops.
Then, the cloud OODA loop integrates the regional information from  edge layers to form a global understanding of the  situation, informing strategic decisions about frequency usage and allocation. 
These decisions are then transmitted back down to the edge and terminal layers, ensuring a unified and coordinated view of the operating environment and enabling seamless adaptation to environment variations.

The implementation of OODA stages is tailored to each layer's operational requirements. The terminal-layer processing prioritizes real-time execution, with efficient detection algorithms analyzing the raw sensor data instantly, local navigation  driving rapid trajectory adjustments, and flight control systems executing time-critical safety decisions. The edge-layer operations center on regional coordination, aggregating terminal data to construct comprehensive situation maps that support coordinated maneuvers and optimized resource allocation. At the cloud layer, the global data integration enables strategic mission planning through advanced algorithms, with subsequent parameter updated widely in the network. Such tiered approach maintains appropriate processing granularity at each level while preserving the consistency of the  system.

By integrating the autonomous operation of UAV swarms with the nested CET H-OODA framework, we can achieve a more robust and adaptive solution for real-time decision-making in complex electromagnetic environments.  The nested OODA loops collectively create a multi-scale decision-making system that enhances both situational awareness and operational adaptability. At the finest scale, the terminal-layer loops provide real-time micro-level awareness of immediate environments through direct sensor inputs, enabling rapid local responses. These local observations aggregate at edge layers to form meso-level regional awareness, supporting coordinated adaptations across UAV subgroups. The cloud layer synthesizes this information into the macro-level mission awareness for the strategic adaptation. The continuous bi-directional information flows between layers ensure local actions remain mission-aligned while global strategies adapt to ground realities.
The UAV swarms can adjust their transmission parameters in real-time to optimize frequency usage, ensuring efficient communications and low interferences. 

\begin{figure*}[t]
\centering
\includegraphics[scale=0.6]{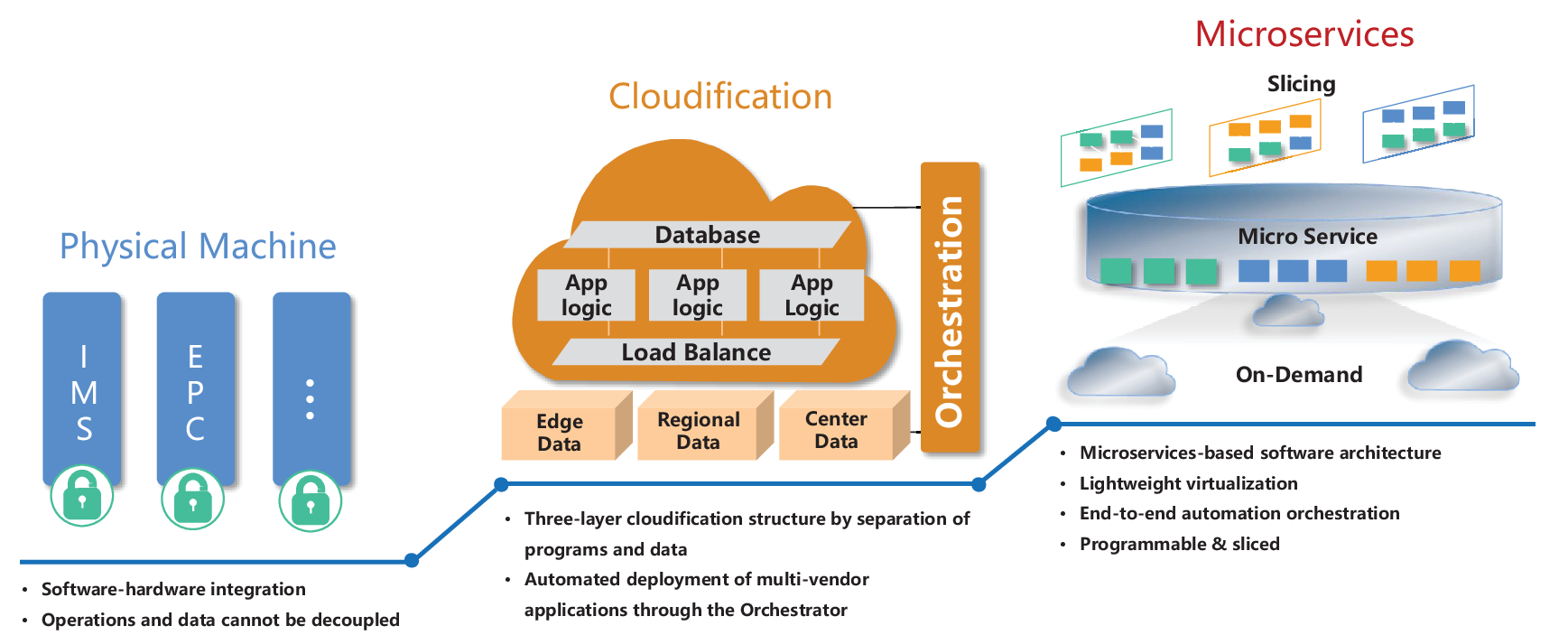}
\caption{\label{fig:VNF} Network cloudification evolution of NFV.}
\end{figure*}

\section{Integration of NFV with H-OODA in UAV Swarms}

The implementation of H-OODA loops in UAV swarms necessitates flexible configurations of network functions and adaptive mechanisms, while the NFV techniques can  provide flexible and scalable decision-making functions. Therefore, the integration of NFV and  H-OODA  brings more insights for the UAV swarm operation.

\subsection{Preliminaries of NFV}

NFV can enable the virtualization of network functions, enabling the deployment and management in a flexible and scalable manner.  
In detail, Fig. \ref{fig:VNF} illustrates the cloudification evolution of NFV, which  transforms the traditional physical machines into a more flexible and scalable microservices-based architecture. 
In particular, the traditional form is characterized by tightly integrated components, such as Internet protocol multimedia subsystem (IMS) and evolved packet core (EPC), which are heavily dependent on hardwares. 
In contrast, the cloudification involves the separation of programs and data, as well as the automated deployment of multi-vendor applications via the orchestrator. 
By decoupling the network functions from the underlying hardwares, NFV can facilitate the  dynamic computing resource allocation, scalable communication mechanism, and rapid network function reconfiguration.

Moreover, the virtualization of network functions enables the creation of complex network services composed of multiple functions. 
Specifically,  the NFV-based service function chain (SFC) can be leveraged to create a flexible and scalable service function path \cite{wu2024adaptive} that supports H-OODA implementation in UAV swarms.
By chaining  multiple network functions such as data collection, data analysis, and decision-making, SFC can enable the H-OODA of UAV swarms to collect and analyze data, make decisions, and take actions in a rapid and efficient manner.

\subsection{Applying NFV to Enhance H-OODA in UAV Swarms}

The NFV enabled H-OODA mechanism presents a transformative approach to optimize UAV operations in dynamic environments. 
In particular, UAV swarms can leverage the virtualized network functions,  programmable network topologies, and resilient resource optimizations,  thereby enhancing the operational effectiveness. 
The integration also enables UAV swarms to simultaneously execute multiple missions, instantly respond to unexpected circumstances as well as  requirements, and maintain comprehensive  situational awareness of dynamic environments.
In detail, NFV plays a crucial role at each stage of the H-OODA framework as follows. 

\begin{itemize} 
\item In the \textit{Observe} stage, NFV facilitates the rapid deployment and scaling of virtualized data collection modules, ensuring that UAVs can swiftly gather relevant information without being constrained by fixed infrastructure limitations.

\item  In the \textit{Orient} stage, NFV supports the analysis and interpretation of the collected data through SFC, which enables the dynamic chaining of multiple virtual network functions (VNFs) for data filtering, aggregation, and preliminary analysis. The orchestration of SFCs enables flexible data routing, ensuring the important information can be processed quickly and accurately.

\item In the \textit{Decide} stage, the orchestration capabilities of NFV provide adequate resources and facilitate the execution of advanced decision-making algorithms \cite{oubbati2023multi}, enabling UAV swarms to make informed and data-driven decisions.
Besides, when VNFs are equipped with predictive analytics and  intelligence abilities, UAV swarms can deal with more complex missions  efficiently. 

\item  In the \textit{Act} stage, NFV leverages the software-defined networking (SDN) technique to establish virtualized, mission-specific communication networks \cite{SDN_UAV2024}, ensuring reliable and timely data exchange between UAV swarms and ground control stations. The centralized control plane of SDN can optimize network traffic and dynamically adjust network configurations in response to changing conditions, thereby enhancing the overall responsiveness of UAV operations. 
\end{itemize} 

The H-OODA framework demonstrates clear advantages over traditional OODA implementations through its hierarchical architecture. By distributing processing across multiple layers, the framework achieves faster response time while maintaining effective coordination through edge-layer aggregation. Furthermore, the integration of NFV technology enables superior adaptation capabilities.

\section{Case Studies and Experimental Results}

To validate the effectiveness of the proposed H-OODA framework, we investigate the performance in a UAV swarm based target search scenario for the OODA loops of single-layer, edge-end, and H-OODA.
The experiment is conducted in a simulated environment with a fixed 100$ \times $100 square meters area, where 15 UAVs are deployed to search for moving targets. 
Each UAV is equipped with sensors that have a specific sensing range, and they utilize the OODA loop to search and track the target within this range. 
To ensure the results reliability, the experiment is repeated 10,000 times and the average results are obtained.

\begin{figure}
\centering
\includegraphics[scale=0.34]{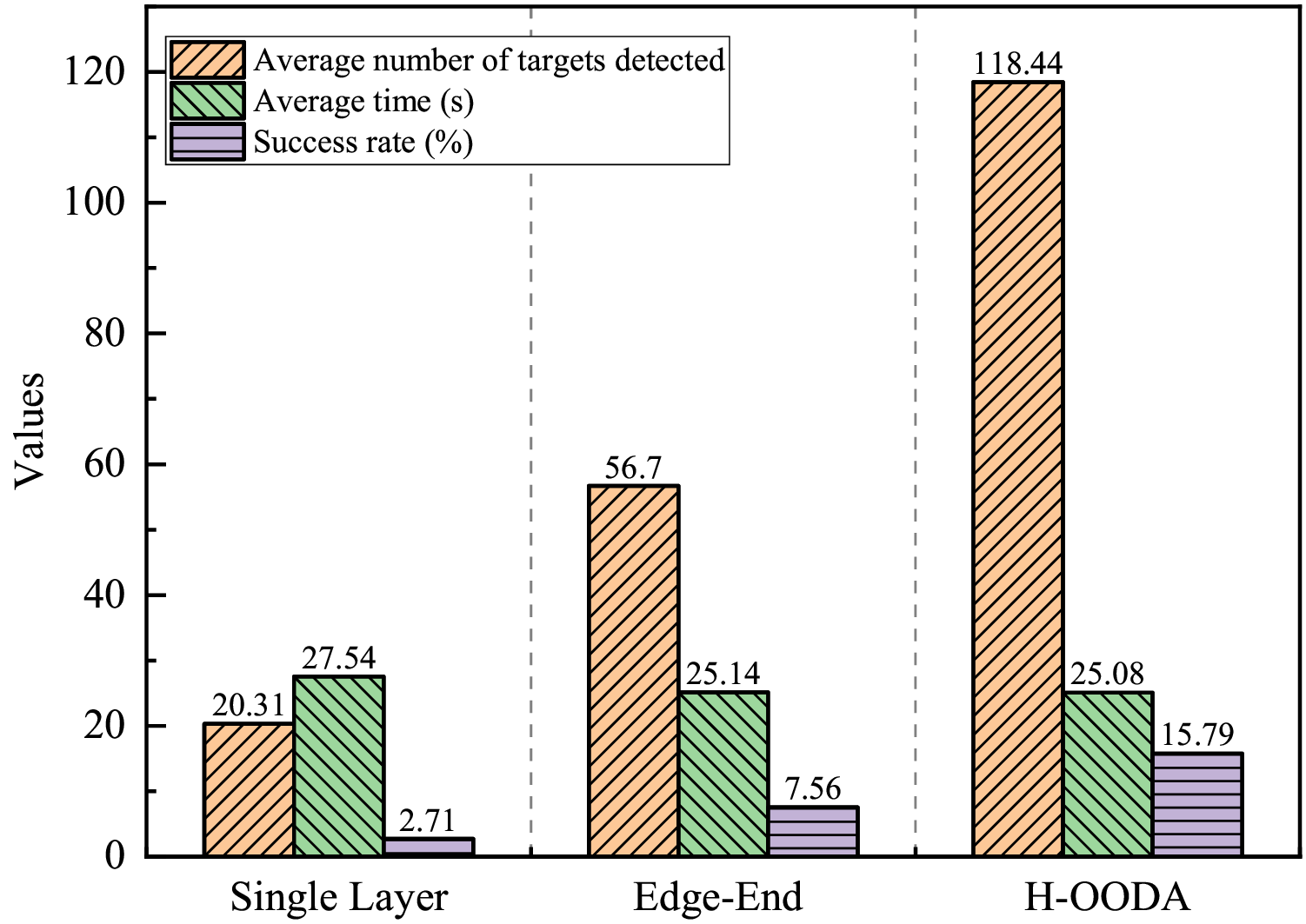}\caption{Performance comparison under different OODA frameworks.}
\label{search}
\end{figure}

As shown in Fig. \ref{search}, the results demonstrate that the H-OODA framework outperforms both the single-layer and edge-end OODA framework in terms of search efficiency, target search time, and success rate.
The  H-OODA framework is adept at searching and tracking moving targets, achieving higher search efficiency and success rates, which is accounted by that  the H-OODA framework leverages  the advantages of each layer and  enhances the overall performance. 
In particular, the cloud layer provides a broader environmental perspective, while the edge and end layers offer more detailed information about the target location. 
The combination of multiple layers enables the H-OODA framework to make informed decisions and adapt effectively to situation variations. 
In contrast, the single-layer OODA framework relies solely on its sensing range, limiting its ability to search and track targets. Although the edge-end OODA  outperforms the single-layer OODA, it still exhibits limitations in sensing range and decision-making capabilities.

\begin{figure}[ht]
\centering
\subfloat[QoE under different H-OODA loops.]{%
    \includegraphics[scale=0.34]{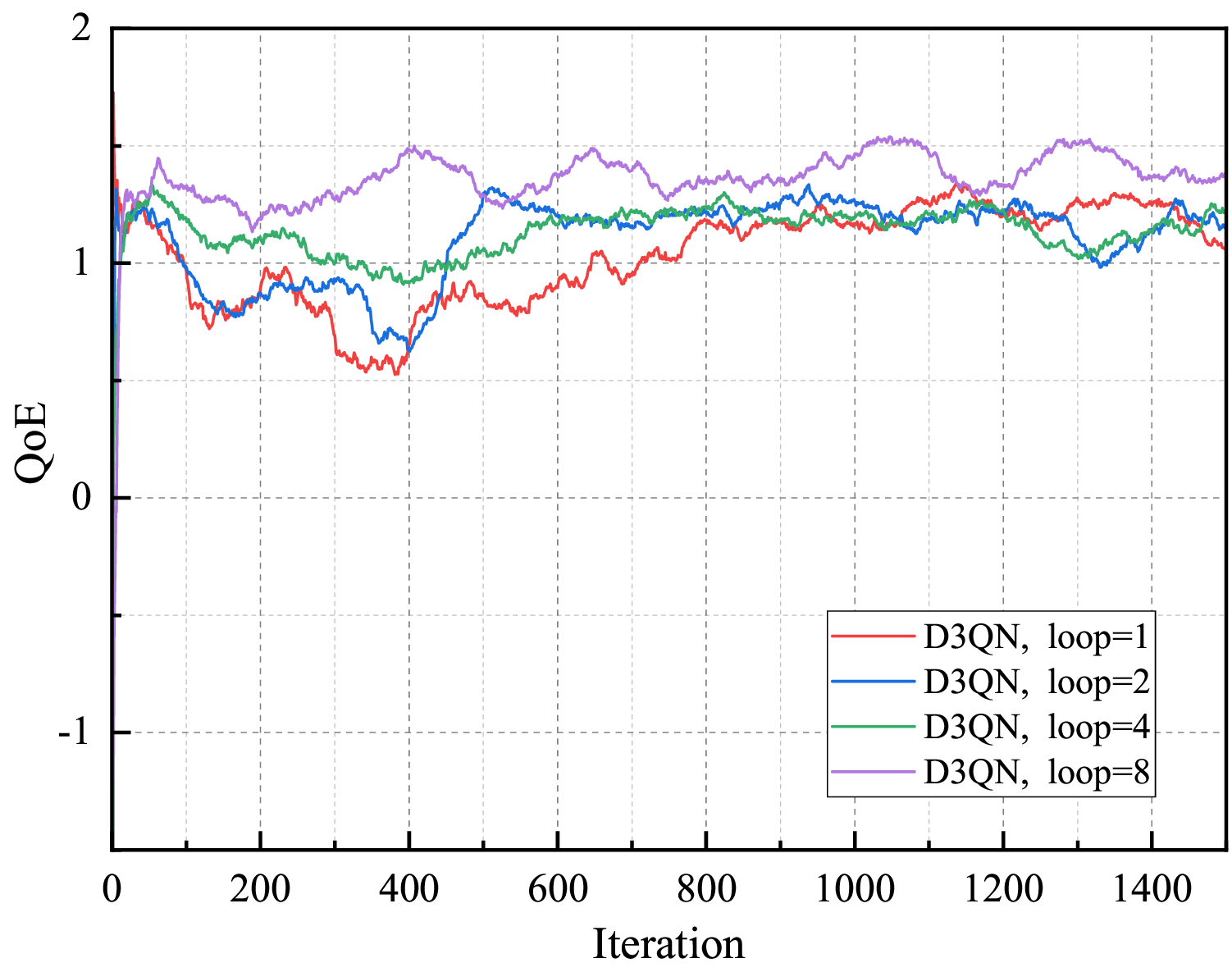}%
    \label{Qoe}%
}
\\
\subfloat[Error rates under different H-OODA loops.]{%
    \includegraphics[scale=0.34]{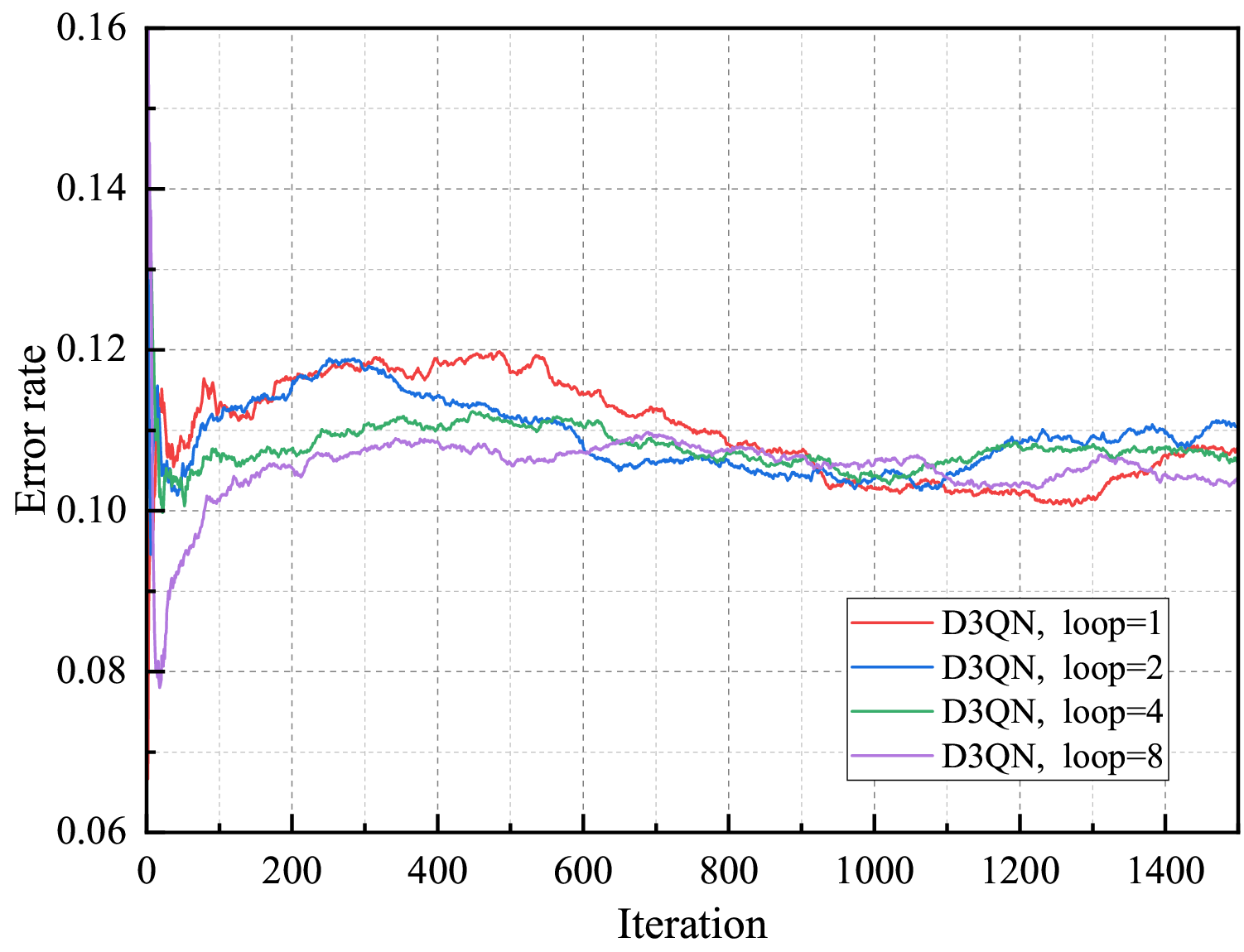}%
    \label{Error}%
}%
\caption{Comparisons of different H-OODA loops.}
\label{fig:both}
\end{figure}

To further assess the benefits of NFV enabled H-OODA framework, we analyze two key performance metrics of the system service quality of experience (QoE) and  error rate of SFC deployments, in which QoE is related with delay, energy cost, as well as bandwidth allocation.
Besides, the impact of the H-OODA depths, is defined as the number of cycles (1, 2, 4, and 8) that the loop executes  on these metrics, realized by the dueling double deep Q-network (D3QN) algorithm.
As depicted in Fig. \ref{Qoe}, the results show that  QoE of the system improves significantly with increasing H-OODA loop depth, since deeper H-OODA loops  consider a wider range of situations and possibilities, enabling the system to make more informed decisions and take more accurate actions. 
Consequently, at the loop depth of 8, the system achieves superior user experience, demonstrating a positive correlation between H-OODA loop depth and QoE.
Moreover, as illustrated in Fig. \ref{Error}, the error rate of the system exhibits a downward trend with increasing depth of H-OODA loop, since deeper OODA loops can more accurately identify and respond to various situations, effectively mitigating erroneous decisions and actions. 
As a result, at the loop depth of 8, the system demonstrates enhanced reliability and stability, highlighting the critical role of deepened OODA loops in ensuring robust system performance.

The experimental results show that the H-OODA framework achieves reliable target detection and mission completion under varying conditions.  The hierarchical coordination among cloud, edge, and terminal layers ensures efficient operations in dynamic environments and rapid adaptations to changing scenarios, making the framework well-suited for applications such as disaster response and urban surveillance. Its distributed architecture and dynamic resource allocation further ensure the stable performance across operational scales.

\section{Challenges and Future  Directions} 
The design of the H-OODA framework for UAV swarms holds significant promise to enhance the  decision-making and adaptability in dynamic environments. However, the following  challenges should still be addressed, which are summarized  in Table \ref{table}, including the data processing complexity, communication reliability and bandwidth limitations,  balance of autonomy and human supervision, as well as  security and resilience. 
Moreover, we explore the possible  directions to deal with these challenges for future researches. 

\subsection{Challenges of Implementing H-OODA in UAV Swarms} 
\subsubsection{Data Processing Complexity}
A primary challenge to deploy  the H-OODA loop in UAV swarms is the large volume and complexity of data generated by multiple sensors, cameras, etc. It requires advanced data analyses,  machine learning algorithms, and distributed computing resources to swiftly process the data, transform it into executable information, and make corresponding timely decisions.

\subsubsection{Communication Reliability and Bandwidth Limitations} 
The reliable and low-latency communication is vital for real-time decision-making and coordination in UAV swarm operations.
However, maintaining seamless connections of individual drones  with the edge and cloud layers is  challenging, since the limitations of bandwidth, signal interference, and potential  disruptions in unpredictable environments can all hinder reliable communication. Therefore, robust communication mechanisms are essential for adapting to dynamic conditions.

\subsubsection{Balance of Autonomy and Human Supervision} 
It is challenging to strike  a balance between the autonomy degree  of UAV swarms and the need for human supervision. While increasing autonomy can enhance the  agility and responsiveness of UAV swarms, it also raises concerns regarding accountability and ethical decision-making. Hence, the well-defined protocols, clear lines of responsibility, and ethical considerations in mission planning are essential for achieving this balance.

\subsubsection{Security and Resilience} 
UAV swarms are inherently susceptible to a variety of security threats, including cyberattacks, signal jamming, and physical interference. A security breach may result in disastrous consequences,  jeopardizing the safety and integrity of swarm operations and their services. Consequently, the development and implementation of robust cybersecurity measures and resilience strategies are crucial to mitigating these threats and ensuring the reliability of UAV swarm operations.

In brief, the implementation of the H-OODA loop in UAV swarms poses new challenges. To ensure successful and responsible operations, it is essential to design  holistic approaches that addresses the interdependencies among the  data processing, communication, autonomy, and security. By tackling these challenges, we can fully leverage the potentials of H-OODA enabled UAV swarms in dynamic environments.

\setlength{\extrarowheight}{3pt}
\begin{table}[t]
\centering
\caption{\label{table}Challenges of H-OODA in UAV Swarms}
\begin{tabular}{|>{\centering\arraybackslash}m{0.38\linewidth}|>{\centering\arraybackslash}m{0.48\linewidth}|}
\hline
\textbf{Challenge} & \textbf{Research Direction} \\
\hline
\multirow{3}{=}{\makecell{Data processing complexity}} & - Algorithm optimization\\
& - Advanced edge computing\\
& - Distributed data fusion\\
\hline
\multirow{4}{=}{\makecell{Communication reliability and\\bandwidth limitations}} & - Enhanced communication protocol\\
& - Dynamic spectrum management\\
& - Multi-modal communication\\
& - Cognitive radio technology\\
\hline
\multirow{3}{=}{\makecell{Balance of autonomy and\\human supervision}} & - Explainable AI \\
& - Human-machine collaboration\\
& - Ethical decision framework\\
\hline
\multirow{3}{=}{\makecell{Security and resilience}} & - Post-quantum cryptography\\
& - AI-enhanced threat detection\\
& - Resilience testing protocol\\
\hline
\end{tabular}
\label{tab:research_directions}
\end{table}

\subsection{Future  Directions}
\begin{itemize} 
\item To deal with the data processing complexity, the developments of advanced algorithmic optimization techniques, edge computing mechanisms, and distributed data fusion methodologies are crucial. These directions can enable the efficient processing of vast datasets, ensuring swift and accurate decision-making in dynamic  environments. On this basis,  UAV swarms can effectively process and analyze large amounts of data in real-time, thereby facilitating informed decision-making.

\item To address the communication reliability and bandwidth limitations, the directions  to develop enhanced communication protocols, dynamic spectrum management strategies, multi-modal communication systems, and cognitive radio technologies are significant. These innovations can ensure robust and low-latency links, even with the signal interference and potential disruptions, thereby enabling the seamless operation of UAV swarms in unpredictable terrains.

\item To balance the autonomy with human supervision, it is essential to establish well-defined rules for autonomy that incorporate explainable AI, human-machine collaboration, and ethical decision mechanisms. By designing protocols that ensure clear lines of responsibility and foster collaborations between humans and UAV swarms, we can leverage the strengths of both human intuition and machine efficiency. These approaches can improve the trust and accountability for autonomous decision-making of UAV swarms.

\item To guarantee the  security and resilience for UAV swarm operations,  the development of post-quantum cryptography, AI-enhanced threat detection, and resilience testing protocols is crucial. Future works can explore integrating machine learning techniques to enhance the environmental perception and threat assessment capabilities within the H-OODA framework. The development of hybrid decision-making models may further  balance the operational reliability and adaptive performance. By integrating these technologies, it strengthens the cybersecurity measures  and helps develop strategies to mitigate potential threats. Such an integrated approach provides a robust defense against cyber threats and ensures the reliable operation of UAV swarms.
\end{itemize} 

In short, the H-OODA loop facilitates UAV swarms with the potentials to revolutionize operations in uncertain environments, but also brings non-negligible challenges.
By addressing  these challenges and exploring possible future directions, the H-OODA enabled UAV swarms can realize safe and effective deployment and implementation.

\section{Conclusions}
In this work, we have proposed the H-OODA loop framework to address the efficient task implementation for UAV swarms in uncertain environments. 
The mechanism of H-OODA has been particularly analyzed  to reveal the insights in UAV swarms.
Moreover, we have investigated the potentials of combining H-OODA with NFV to improve the flexibility and efficiency of UAV swarms.  
Besides,  detailed use cases have been conducted to validate the advantages of H-OODA. 
In addition, the challenges and possible directions of H-OODA framework have been analyzed.  The multi-layer H-OODA framework  enables seamless scalabilities from simple surveillances to complex disaster response operations, maintaining operational effectiveness through adaptive coordination between global mission control and local decision-making. Such a scalable architecture fulfills  essential requirements in UAV swarm applications with dynamically changing  demands.
We envision that future research will continue to address the open challenges of H-OODA in UAV swarms.

\bibliographystyle{IEEEtran}

\bibliography{wu2022unmanned,alharasees2023human,alharasees2023evaluating,fan2021mission,mccoy2019software,
zhu2017guide,messaoudi2023survey,xiao2019nfvdeep,sun2023effectiveness,qu2023elastic,chen2023improving,zhuang2019sdn,yang2023depth, zirui2024TON, 
fu2019dynamic,li2019virtual,shu2020novel,wu2024adaptive,PERA_OODA2024, quantumC, NFV_aerial, SDN_UAV2024, OODA_Boyd2020,harchaoui20245g,motlagh2017uav,ouahouah2021deep,bekkouche2020service,bouzid20235g,ameur2024efficient,oubbati2023multi}

\end{document}